\documentclass{jltp}

\title{The spectral conductance of a proximity superconductor and the 
re-entrance effect}
\author{H. Courtois,$^1$ P. Charlat,$^1$ Ph. Gandit,$^1$ D. 
Mailly,$^2$\\ and B. Pannetier$^1$}
\address{$^1$ C.R.T.B.T.-C.N.R.S., in association with Universit\'e 
Joseph Fourier,\\ 25 Av. des Martyrs, B.P. 166, 38042 Grenoble Cedex, France\\
$^2$ L.M.M.-C.N.R.S., 196 Av. H. Ravera, 92220 Bagneux, France}
\runninghead{H. Courtois {\it et al.}}{The spectral conductance and the 
re-entrance effect}

\input epsf
\epsfverbosetrue
\begin{document}

\maketitle
\begin{abstract}
In a mesoscopic metal in proximity with a superconductor, the 
electronic conductance is enhanced in a very energy-sensitive way. In 
this paper, we discuss the spectral conductance of a proximity 
superconductor from both the theoretical and experimental point of 
view. The dependence of the spectral conductance on the 
phase-breaking length, 
gap of the superconductor and interface transparency is theoretically 
investigated. We present experimental data on the re-entrance of the 
normal-state conductance at very low temperature and bias voltage. 
A complete description of the experimental 
data needs taking into account heating of the 
reservoirs by the bias current. In addition, we show that the energy 
sensitivity of the proximity effect enables one to access the energy 
distribution of the conduction electrons inside a mesoscopic sample.
\end{abstract}

\section{INTRODUCTION}

\subsection{The general context}

In the past few years, the study of the proximity effect in hybrid 
structures made of a normal metal in contact with a superconductor 
has known a remarkable revival.\cite{Revue_Lambert,Comments,Esteve} This 
originated from both the availability of a new generation of tailored-made 
submicron-sized samples and the new understanding 
of the coherence effects in electron transport at the mesoscopic scale.
The proximity effect is the generic name for the phenomena appearing 
at the interface of a normal metal (N) with a superconductor (S). In a 
N-S geometry, the N-metal shows superconducting-like properties 
including a magnetic screening,\cite{Mota} a modification of 
the density of states\cite{Gueron} and an energy-dependent conductance 
enhancement.\cite{Petrashov,CourtoisPrl,Taka} In a S-N-S geometry, 
the Josephson effect is the manifestation of the coherence in the 
proximity effect. Eventually, the conductance of a N-I-S tunnel junction is 
governed by the tunneling of electron pairs.

In a N-I-S tunnel junction, the BTK theory\cite{BTK} describes the 
cross-over in the  behaviour of the interface conductance as a 
function of tunnel barrier transparency. The interface conductance 
is twice the normal-state one in the high-transparency 
limit ($G_{NS} = 2G_{NN}$) while it vanishes ($G_{NS} \rightarrow 0$) 
in the low-transparency limit. The discovery of the zero-bias anomaly\cite{Kastalsky} showed a clear 
discrepancy with the BTK theory : in a confined geometry the 
conductance of the N-I-S junction does not go to zero at zero temperature but 
increases at low temperature and bias. The physical picture for this 
effect is that the confinement of electrons by the disorder in the 
vicinity of the N-S interface makes the transmission probabilities add 
coherently. This enhances drastically the tunneling of 
pairs.\cite{vanWees,HekkingNazarov}

Diffusive electron transport in the N metal part of a N-S junction 
is also significantly enhanced by the proximity 
effect.\cite{Petrashov,CourtoisPrl,Taka}
The observed large amplitude of the conductance enhancement used to 
appear in contradiction with the prediction of the Bogoliubov-de 
Gennes equations\cite{Lambert} and the random matrix 
theory.\cite{Beenakker} The observation of the re-entrance effect has 
been the key experiment that made the connection between the 
high-temperature regime and the low-temperature 
regime.\cite{PrlCharlat,Hartog,Petr_1998,Poirier} At finite temperature or 
bias voltage, the conductance of the "proximity superconductor" made 
of the N metal in proximity with S is significantly enhanced, while it 
should coincide with the normal-state value at zero temperature and bias. 
These features arise clearly from the quasiclassical theory.

In this paper, we will focus on the conductance enhancement in 
very small N-S structures. We will consider the regime where the N-S 
interface conductance is large compared to the metallic 
conductance of the N metal wire. The small size of the structures enables 
us to investigate the transport properties at temperatures well below 
the characteristic energy of the proximity effect.

In the following, we will first introduce the main theoretical concepts of 
the proximity effect. We will concentrate on the spectral conductance 
and its dependence on various physical parameters. Afterwards, we will 
present a thorough experimental study of the re-entrance effect in a 
mesoscopic structure made of a normal metal wire in contact with a 
superconductor. We will compare our experimental results with the prediction 
of the theory and conclude about the energy distribution in the reservoirs. 
Eventually, we will show that the large energy-sensitivity of the proximity 
effect can be used to test the electron energy distribution inside a 
mesoscopic sample.

\subsection{The Andreev reflection}

\begin{figure}
\epsfxsize=12 cm
\epsfbox{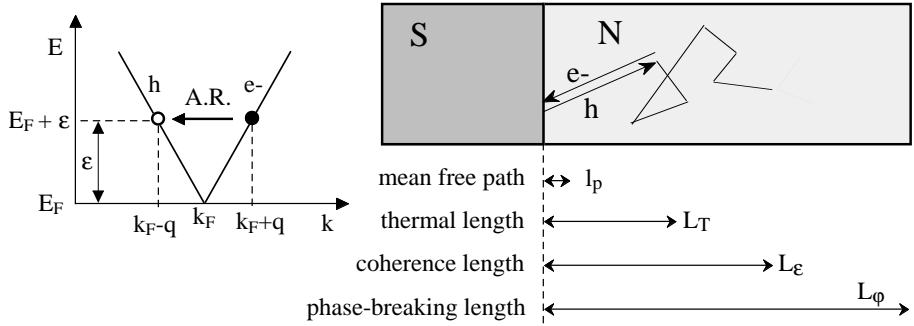}
\caption{Left : Schematic of the Andreev reflection process. An 
incident electron with an extra energy $\epsilon$ compared to the 
Fermi energy $E_F$ hits the N-S interface from the N side. The 
reflected hole-like particle and the incident electron have a slight 
wave-vector mismatch $2q$. Right : Relevant length scales with their 
schematic respective amplitudes in a metallic thin film. The 
energy-dependent coherence length $L_\epsilon$ is the length over 
which the two components of the Andreev pair acquires a phase difference 
of order $\pi$.}
\label{Andreev}
\end{figure}

Let us consider the interface between a normal metal and a 
superconductor. Well below the superconducting transition temperature of S and at 
low bias voltage, the thermal energy $k_B T$ and the electrostatic 
energy $eV$ are much smaller than the gap $\Delta$ of S. In this regime,
single electrons cannot enter the superconductor because of the absence of 
avalaible states at the same energy. As a consequence, 
electrons arriving from N will be Andreev-reflected at the N-S 
interface, see Fig.\ref{Andreev}. In this process, the electron is 
retro-reflected as a hole, while a Cooper pair is transmitted in the 
superconductor.\cite{Andreev}

At the Fermi energy, the reflected hole has the opposite spin, the opposite 
velocity and the same momentum as the incident electron. The reflected hole traces back 
the trajectory of the incident electron. The reflection of a hole may 
also be seen as the absorption of a second electron. As the Andreev 
reflection correlates the two electron states, it is 
equivalent to the diffusion of pairs in the N metal. We call the 
diffusing pairs in N "Andreev pairs" as their existence is not due to an 
intrinsic attractive interaction in N, but to a remote effect which is 
the Andreev process at the N-S interface.

If one looks more precisely, the reversal of every electron velocity 
component is perfect only at the Fermi level.\cite{BTK} Let us consider an electron 
with a small extra energy $\epsilon$ compared to the Fermi level 
energy $E_F$. The electron wave-vector $k_e = k_F + q$ is larger than 
the Fermi wave-vector $k_F$. The reflected hole has a slightly different 
wave-vector $k_h = k_F - q$, see Fig. \ref{Andreev}. This means that 
the incident and reflected particles will have a difference
\begin{equation}
\delta k = 2q = k_F  \frac{\epsilon}{ E_F}.
\label{deltak}
\end{equation}
in wave-vector. This difference concerns the wave-vector component 
which is perpendicular to the N-S interface. With an arbitrary angle 
of incidence, the change in wave-vector will change the trajectory angle with 
respect of the N-S interface.\cite{Blom} In brief, the 
retro-reflection is imperfect at finite energy.

\subsection{The diffusion of "Andreev pairs"}

In the following, we will focus on the case of a mesoscopic normal 
metal in the dirty limit.  This means that the elastic mean free path $l_{e}$
is much smaller than the sample length $L$ which is itself smaller than 
the phase-breaking length $L_{\varphi}$ : $ l_{e} \ll L \ll L_{\varphi}$.

Let us consider the trajectories of an electron incoming on the 
N-S interface and of the reflected hole nearly retracing the 
trajectory of the incident electron. At zeroth order, the phase 
acquired by the electron is eaten up as the hole retraces back the 
trajectory of the incident electron. Looking a little closer, the 
wave-vector mismatch between the electron and the hole makes the 
phase difference between the hole and the electron 
increase monotonically. After diffusion over a distance $L$ from the 
interface, the phase shift between the two particles is of order $\pi$
at a distance equal to the energy-dependent coherence length :
\begin{equation}
L_\epsilon = \sqrt{\frac{\hbar D}{ \epsilon}},
\end{equation}
$D$ being the diffusion coefficient in N. At the 
same time, the trajectories of the electron and the hole are shifted 
by a distance of order the Fermi wavelength.\cite{Blom} The exact value of the 
spatial shift depends on the incidence angle on the N-S interface. 
Further diffusion of the two particles will therefore be different 
and the pair will break apart. In this respect, the length $L_{\epsilon}$ 
actually plays the role of the coherence length of the Andreev pair.

The coherence length $L_\epsilon$ coincides 
with the thermal length\cite{de_Gennes} $L_T$ at the energy $ \epsilon = 2\pi k_B T$. 
The thermal length $L_T$ is relevant as soon as one considers the 
whole electron distribution at thermal equilibrium. This is for 
instance the case in the Josephson effect. At the Fermi 
level, the length $L_\epsilon$ diverges. Indeed, the coherence 
length of an Andreev pair is limited by the phase-breaking 
length $L_{\varphi}$ of a single electron. The average coherence 
length of an Andreev pair therefore varies from about the thermal coherence 
length $L_T$ at high energy ($\epsilon \simeq k_{B}T$) to the phase-coherence 
length $L_\varphi$ at low energy ($\epsilon \simeq 0$).

The correlation between the electron and the reflected hole can also 
be described in terms of energy with the equivalence :
\begin{equation}
L_{\epsilon} = L \Leftrightarrow \epsilon = \epsilon_{c}
\end{equation}
where
\begin{equation}
\epsilon_c = \frac{\hbar D}{ L^2}
\end{equation}
is the Thouless energy, or correlation energy, of a sample of length $L$. The Thouless energy 
is the fundamental energy scale for the proximity effect at the spatial 
scale $L$.\cite{CourtoisPrl} At a given distance $L$, only electrons with 
an energy below the Thouless energy are still correlated as Andreev pairs.
The other Andreev pairs are broken.

\section{THE THEORETICAL BACKGROUND}

\subsection{The Usadel equation}

The diffusion of superconductivity in an inhomogenous structure can 
be described by the Gorkov Green's functions. In the limit where all 
the experimentally relevant length scales are much larger than the 
Fermi wavelength, the simplification of this fully quantum theory 
into the quasiclassical theory is 
valid.\cite{Larkin,Volkov,Zhou,Nazarov96,Wilhelm,Yip,Volkov-Lambert} 
The Usadel equations are obtained in the diffusive regime where the 
elastic mean free path is small. In this framework, weak 
localization effects and conductance fluctuations are neglected.

Let us restrict to the case of a one-dimensional N-S structure in 
zero magnetic field and no superconducting phase gradient. The latter 
condition is fulfilled as soon as there is only one superconducting 
island. In the absence of electron-electron interaction in N, 
the Usadel equation in the N metal is :
\begin{equation}
\hbar D \, \delta_x^2 \theta+\{\,2i\epsilon-\frac{\hbar 
D}{L_\varphi^2}\,\cos\theta\}\sin\theta=0
\label{Usadel}
\end{equation}

The complex proximity angle $\theta = \theta_{1} + i\theta_{2}$ is related 
to the anomalous Green function $F$ :
\begin{equation}
F(\epsilon,x)=-i  \sin[\theta(\epsilon,x)]
\end{equation}
The anomalous Green function $F$ is often called the pair amplitude 
although if differs from the actual pair density which takes into account the  
energy distribution of the electrons.

The complex angle $\theta$ and the pair amplitude $F$ are functions of 
both the energy above the Fermi level $\epsilon$ and the distance 
from the S interface $x$. The Usadel equation features a non-linear 
diffusion equation for the proximity angle $\theta$. Let us note 
that the phase-breaking length $L_{\varphi}$ enters the Usadel 
equations as a cut-off for the proximity effect. The inelastic mean 
free path $L_{in}$ is included in $L_{\varphi}$, but does not show up by 
itself. It enters directly in the problem only through the energy 
redistribution of the electrons in N.

At the contact with an N metal reservoir, the angle $\theta$ 
is zero. At the contact with a superconducting island S, a complete 
treatment implies solving the self-consistency equation for the pair 
potential in the vicinity of the interface.\cite{Sols,Bagwell} Here 
we will restrict to the idealized case of a step-like pair 
potential. The pair potential will be considered as being 
constant and equal to the gap in S, and zero in N. With this 
assumption, the boundary condition for a perfectly transparent N-S interface 
is $\theta = i \pi/2$ at energy zero or well below the gap $\Delta$.

\subsection{The spectral conductance}

Electron transport is non-local at the mesoscopic scale. 
In the framework of the quasiclassical theory, the electron 
transport is nevertheless determined by a local conductivity 
$\sigma(\epsilon,x)$  which expresses as :
\begin{equation}
\sigma(\epsilon,x)=\sigma_N \, \cosh^2\theta_{2}(\epsilon,x)
\end{equation}
where $\sigma_N$ is the normal-state conductivity. From this relation, 
the local conductivity is always larger than the normal-state conductivity. 
Since the function $\theta_{2}$ is strongly energy-dependent, so is the 
conductivity enhancement. 

To proceed with transport properties, one needs to know the occupation of 
current-carrying states. The non-equilibrium quasiparticle energy 
distribution function can be derived from the out-of-equilibrium 
Keldysh Green function.\cite{Larkin} The generalized quasiparticle 
distribution function has two components : an odd part and an even 
part. The odd part is related to the energy distribution functions 
of the electrons $f_{e}$ and the holes $f_{h}$ as : 
$f_{{\it 0}}(\epsilon)=1/2(f_{e}+f_{h})$, and reduces to 
$\tanh(\epsilon/2k_{B}T)$ at equilibrium. The even part $f=1/2(f_{e}-f_{h})$ 
is related to the imbalance in the population of holes and 
electrons, and reduces to zero at equilibrium (no current). It can 
therefore be named the out-of-equilibrium part of the distribution 
function.

At any point in a wire of section $S$, the spectral 
current $i(\epsilon,x)$ is related to the spatial derivative of the 
distribution function $f$. In this work, we will consider the regime where 
the inelastic mean free path $L_{in}$ is much larger than the sample length : 
$L_{in}\gg L$. In this regime, a conduction electron keeps its 
energy while travelling through the sample. Therefore, the 
transport channels at different energies are independent and the 
spectral current $i(\epsilon)$ is constant along the wire. This permits 
to write the spectral current as a function of the spectral conductance 
$g(\epsilon)$ and the difference $\Delta f(\epsilon)$ of the 
distribution functions $f$ at the two extremities of the considered wire :
\begin{equation}
i(\epsilon)=\sigma(\epsilon,x) S\, \frac{d f(\epsilon,x)}{d x}=g(\epsilon) 
\Delta f(\epsilon)
\end{equation}
As a consequence, the spectral currents are given by linear circuit 
theory rules where local voltages are replaced by the 
electron distribution functions $f$ at the nodes.\cite{Volkov}
The spectral conductance $g(\epsilon)$ is the average conductance for 
electrons with a given energy $\epsilon$ :
\begin{equation}
g(\epsilon)=S [\, \int^{L}_{0}{\frac{dx}{\sigma}} \, ]^{-1}=\sigma_{N}S 
[\, \int^{L}_{0}{\frac{dx}{\cosh^2\theta_{2}(\epsilon,x)}} \, ]^{-1}=G_N + \delta 
g(\epsilon)
\end{equation}
Here, we do not consider the change in conductance of the opened 
channels with increasing voltage. This was included in the framework 
of the scattering matrix theory by Lesovik et al.\cite{Lesovik}

Let us consider the generic example of a sample made of a quasi-1D 
N wire "n" of length $L$ between a S island and a N reservoir, see 
Fig. \ref{Lin} inset. The spectral conductance is a measurable 
quantity, since it coincides with the zero-temperature differential 
conductance at voltage 
$V=\epsilon/e$ :
\begin{equation}
g(\epsilon)=\frac{dI}{dV}(V=\epsilon/e,T=0)
\end{equation}
The spectral conductance $g(\epsilon)$ is strongly energy-dependent 
and always larger than the normal-state conductance.
In the same way, the excess conductance $\delta g(\epsilon)$ is always positive.
At finite temperature, the differential conductance is the integral of 
the spectral conductance multiplied by the energy derivative of the 
Fermi distribution function :
\begin{equation}
\frac{dI}{dV}(V=\epsilon /e,T)=\int^{- \infty}_{+ \infty}{ 
\frac{d\Delta f}{dV} g(\epsilon) d\epsilon }
\label{Diff_Cond}
\end{equation}
This is equivalent to averaging the spectral conductance 
over an energy window of width $4k_{B}T$. 

Real transport experiments involve more complex circuits than the 
generic two-probe configuration considered above. Let us consider the 
general case of a network of 1D wires connected to several normal 
reservoirs and a unique superconducting electrode. The chemical 
potential of the superconducting island is taken as the reference 
potential ($V=0$). One can then define a spectral conductance matrix 
${\underline {\underline g}}$ for the whole circuit connecting the spectral currents matrix 
${\underline i}$ incoming from the reservoirs with the non-equilibrium distribution 
functions ${\underline f}$ :\cite{These_Pierre}
\begin{equation}
{\underline i}={\underline {\underline g}}.{\underline f}
\label{matrice}
\end{equation}
Integration of this set of relation over the energy gives a direct 
access to the current-voltage characteristic, since the summation of 
the spectral currents leads to the measurable current while integration 
of the $f$ functions of the reservoirs leads to the chemical potential of 
the reservoirs. Let us note 
that a zero total current does not mean that the spectral current 
should be zero. In a reservoir with no injected current, the low-energy 
and high-energy spectral currents are in general different from zero but 
cancel out once integrated over the whole energy spectrum.

\subsection{The re-entrance effect}

\begin{figure} [t]
\epsfxsize=12 cm
\epsfbox{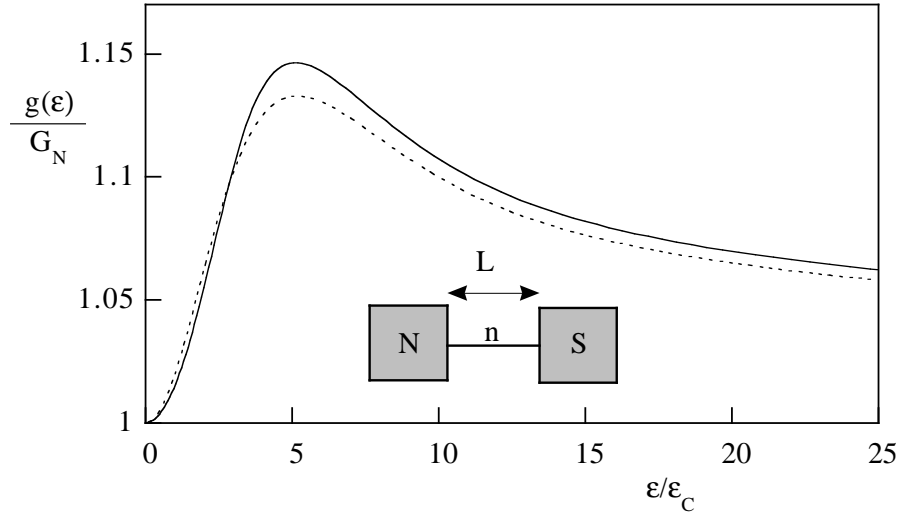}
\caption{Energy dependence of the spectral conductance $g(\epsilon)$ in 
units of the normal-state conductance $G_{N}$ for a N-S sample. The 
sample is made of a normal metal wire "n" of length $L$ between a N 
reservoir and a 
S island, see inset. The spectral conductance was calculated 
with the help of the full non-linear Usadel equations (full line) and in the 
linear approximation (dashed line). The gap and the phase-breaking 
length were assumed to be infinite. The N-S interface transparency is 
considered as equal to 1.}
\label{Lin}
\end{figure}

As an example for the calculation of the spectral conductance, let us 
come back to the generic sample geometry made of an N wire
between an S island and an N reservoir. The numerical solution of the Usadel 
equations for the spectral conductance $g(\epsilon)$ in units 
of the normal-state conductance $G_{N}$ is shown in Fig. 
\ref{Lin}. From the calculation, the spectral conductance shows a 
maximum of about $1.15 \, G_{N}$ at an energy close to 
$5.1\, \epsilon_{c}$ with $\epsilon_{c}= \hbar D/L^2$. 
At higher energy, the spectral conductance decays as $1/\sqrt{\epsilon}$.
The most striking result is that at zero energy, the normal-state 
conductance is recovered. This is the re-entrance effect.

The occurance of the re-entrance of the metallic conductance does 
not depend on the precise sample geometry. It is an exact result from the 
quasiclassical theory,\cite{Volkov,Zhou,Nazarov96,Wilhelm,Yip,Volkov-Lambert} 
the random-matrix theory\cite{Beenakker,Lesovik} 
and the Bogoliubov-de-Gennes equations.\cite{Volkov-Lambert}
The zero-energy conductance of a proximity superconductor coincides 
with the normal-state conductance only in the 
absence of electron-electron interactions. In the presence of 
interactions in the normal metal, the zero-temperature conductance is 
predicted to differ from the normal-state value $G_{N}$. The zero-temperature 
conductance should decrease in the case of repulsive interactions and 
increase in the case of attractive ones.\cite{Nazarov96} The ferromagnetic 
metals provide a case where the interactions are expected to play a great 
role in the behaviour of the proximity effect.\cite{FS}

Let us try to draw a physical picture for the re-entrance effect. The 
conductivity peak for electrons with a given energy $\epsilon$ appears in 
the vicinity of the distance $L_\epsilon$ from the S interface. Strikingly, 
it is the point {\it where the Andreev pairs break}.
At smaller distances $x<L_{\epsilon}$, the Andreev pair is a 
closed object, i.e. the electron and the hole follow exactly the 
same path. Another conduction electron 
cannot enter the Andreev pair. The local conductivity is 
unchanged by the proximity of the superconductor. The density of states is 
nevertheless depleted since the probability to find an isolated electron 
is very small. At a distance $x \simeq L_{\epsilon}$, the electron and the hole 
have distinguishable trajectories. The Andreev state can couple the 
electron with a hole a Fermi wave-length away. The hole can even 
interact again with the superconducting condensate and be 
reemitted as an electron. This `'delocalization" of the electron 
enhances the local conductivity. At high energy, the conductance 
enhancement decays because the electron and the hole are increasingly 
decorrelated.

\subsection{The linearization}

For the easiness of presentation and understanding, it may be convenient 
to linearize the Usadel equation and consider the regime of an 
infinite phase-breaking length. This gives :
\begin{equation}
\hbar D \, \delta_x^2 \theta+2i\epsilon \, \theta=0
\end{equation}
The linearization is fully justified only in the case of a (thin) tunnel 
barrier between N and S. In all cases, it remains a good approximation 
at a distance from the interface larger than $L_{\epsilon}$.

The linearized Usadel equation is a diffusion equation 
for the pair amplitude in the real space dimension $x$. The characteristic 
diffusion length is the coherence length $L_\epsilon = \sqrt{\hbar D 
/ \epsilon}$. At a given position x, the characteristic energy scale 
is $\epsilon_x=\hbar D/x^2$ which coincides with the Thouless energy 
$\epsilon_c=\hbar D/L^2$ in the case $x=L$.

Let us now consider some ideal geometries.

In the case of an semi-infinite length N wire in contact with a S 
island, the angle $\theta$ has the following simple form :
\begin{equation}
\theta=\frac{\pi}{2} \exp[(i-1)\frac{x}{L_\epsilon}]
\end{equation}
The local conductivity can be analytically written as :
\begin{equation}
\sigma(\epsilon,x)=\sigma_N \cosh^2[\frac{\pi}{2} 
\exp(-\frac{x}{L_\epsilon}) \sin\frac{x}{L_\epsilon}]
\end{equation}

In the case of a N wire of finite length, the N reservoir imposes a 
zero value of $\theta$ at the boundary. Fig. \ref{Lin} shows the 
energy dependence of the spectral conductance $g(\epsilon)$ in 
units of the normal-state conductance $G_{N}$. The curves are 
calculated without and with the linearization approximation. We 
observe that the spectral conductance is qualitatively the same in the two 
cases. Although the linearization does not describe accurately the 
contribution of low energies, the 
integrated conductance is very well described. The position and the value of 
the maxima of conductance are very close in the two calculations. 
This comparison is the justification for the linear approximation 
used in Ref. \cite{PrlCharlat}.

\subsection{Influence of the physical parameters}

Let us consider the effect of several physical parameters on the 
spectral conductance of a sample made of a "n" metallic wire between 
one N reservoir and one S island, see Fig. \ref{Lin} inset.

\subsubsection{The phase-breaking length}

\begin{figure}
\epsfxsize=12 cm
\epsfbox{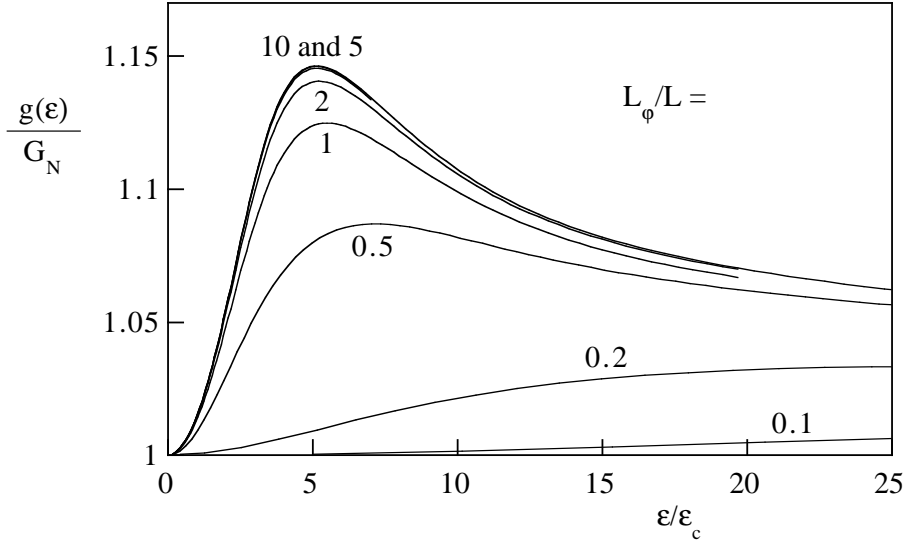}
\caption{Energy dependence of the spectral conductance $g(\epsilon)$ 
in units of the normal-sate conductance $G_{N}$ for various 
values of the phase-breaking length. Each curve is labeled with the 
related ratio of the phase-breaking length $L_\varphi$ over the sample 
length $L$. The gap of S is taken as infinite. The barrier 
transparency is assumed to be $1$.}
\label{Lphi}
\end{figure}

Up to now, we considered that the phase-breaking length $L_\varphi$ is much 
larger than the sample length $L$. In this case, phase-breaking events 
have no effect on the spectral conductance since it is the N reservoir which 
limits the diffusion of pairs in the normal metal wire. Now let us
consider the opposite case where the phase-breaking length is smaller 
than the sample length.

Fig. \ref{Lphi} shows the calculated spectral 
conductance for the Fig. \ref{Lin} sample geometry but with 
a varying value for the phase-breaking length $L_\varphi$.
As soon as $L_\varphi$ becomes of the order of $L$, the 
spectral conductance is affected at energies below 
$\epsilon_\varphi=\hbar D/L_\varphi^2$. When $L_\varphi < L$, the position 
and absolute amplitude of the spectral conductance maximum do not vary 
with the sample length $L$ anymore, but depend directly on the 
phase-breaking length. The spectral conductance maximum shifts to 
$1.2\, \epsilon_\varphi$ with an amplitude equivalent to a $16\, \%$ 
increase for the conductance of the length $L_\varphi$ of the wire.

The phase-breaking brings a very strong cut-off for the 
proximity effect so that it plays the role of an effective sample length. 
As a consequence, the energy $\epsilon_{\varphi}$ also replaces the 
Thouless energy $\epsilon_{c}$.
In a given sample, the phase-coherence length can be shortened 
by the flux induced in the width $w$ of the N wires by an applied 
magnetic field $H$. The effective phase-coherence $L_{\varphi}$ follows the 
relation :\cite{Pannetier-Rammal}
\begin {equation}
L_{\varphi}^{-2}(H)=L_{\varphi}^{-2}(0)+\frac{\pi^{2}}{3} 
\frac{H^{2}w^{2}}{\Phi_{0}^{2}}
\end {equation}
where $\phi_{0}= h/2e$ is the flux quantum and $L_{\varphi}(0)$ is the 
zero-field phase-breaking length. At zero magnetic field, the 
characteristic energy scale of the proximity effect is the minimum of 
the Thouless energy $\epsilon_{c}$ and the phase-breaking--related 
energy $\epsilon_{\varphi}$. As the magnetic field is increased, the 
energy $\epsilon_{\varphi}$ is decreased and the temperature of the 
conductance maximum is shifted to higher temperature. This has been 
observed in a previous experiment on the re-entrance effect.\cite{PrlCharlat}

\subsubsection{The interface transparency}

\begin{figure} [t]
\epsfxsize=12 cm
\epsfbox{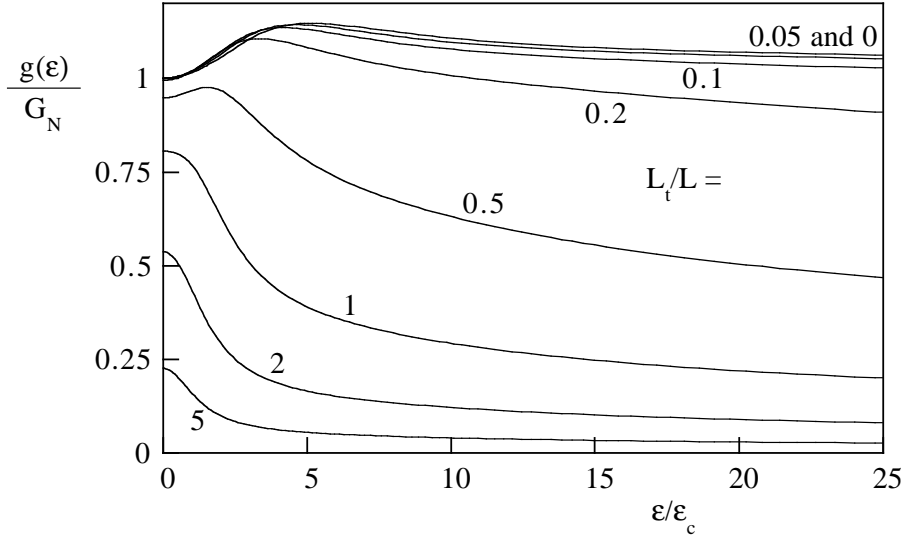}
\caption{Energy dependence of the spectral conductance $g(\epsilon)$ 
for various values of the transparency of the barrier at the N-S interface, in 
units of the normal-state conductance of the N metal alone. The 
calculated conductance takes into account both the metallic 
conductance of the N metal wire and the finite conductance of the 
N-S interface. The curves 
are calculated for different values of the ratio $L_t/L$ of the barrier 
equivalent length $L_t$ over the sample length $L$. The gap of S is 
taken as infinite. The phase-breaking length is considered as infinite.}
\label{Bar}
\end{figure}

In this study, we are mainly interested in the case of a clean, 
metallic, interface between the normal metal and the superconductor. 
Nevertheless, it is worth knowing how much an intermediate interface 
transparency can modify the spectral conductance.
The transparency $t$ of the N-S interface can be smaller than 1 due to 
the presence of a small tunnel barrier at the interface. A mismatch 
between the Fermi velocities of the two metals can also induce a 
finite reflection coefficient at the interface.

The interface transparency can be implemented in the calculation 
as a different boundary condition for the angle $\theta$ at the 
N-S interface :
\begin{equation}
\sin(\theta_{S}-\theta_{N})=L_t  \,\delta_x \theta
\end{equation}
where the length $L_t = l_{e}/t$ is the barrier equivalent 
length,\cite{Zhou} $\theta_{N}$ 
and $\theta_{S}$ are the values of the angle $\theta$ on the N or S 
side of the N-S interface, respectively. The length $L_{t}$ has a 
simple physical explanation in the limit of a small transparency 
since it is the length of normal metal wire which 
has the same resistance as the barrier.

Fig. \ref{Bar} shows the conductance of the total N-S sample, 
including the interface resistance. As already discussed in Ref. 
\cite{Yip}, it shows a cross-over 
between two distinct regimes as the interface transparency is decreased. 
At large transparency we observe a maximum of the spectral conductance 
at finite energy : it is the re-entrance effect. When the 
interface transparency is so small that the length $L_t$ is larger 
than the sample length $L$, the maximum of spectral conductance sits a 
zero energy. This is the zero-bias anomaly of a N-I-S tunnel junction 
conductance. The zero-bias anomaly has been first observed by Kastalsky et
al.\cite{Kastalsky} It can be modulated by a flux in a loop geometry.\cite{Pothier}
The physical interpretation is that the tunneling of pairs through the 
tunnel barrier is enhanced by the confinement by the disorder in the 
N-metal layer.\cite{vanWees,HekkingNazarov,Beenakker}

\begin{figure}
\epsfxsize=12 cm
\epsfbox{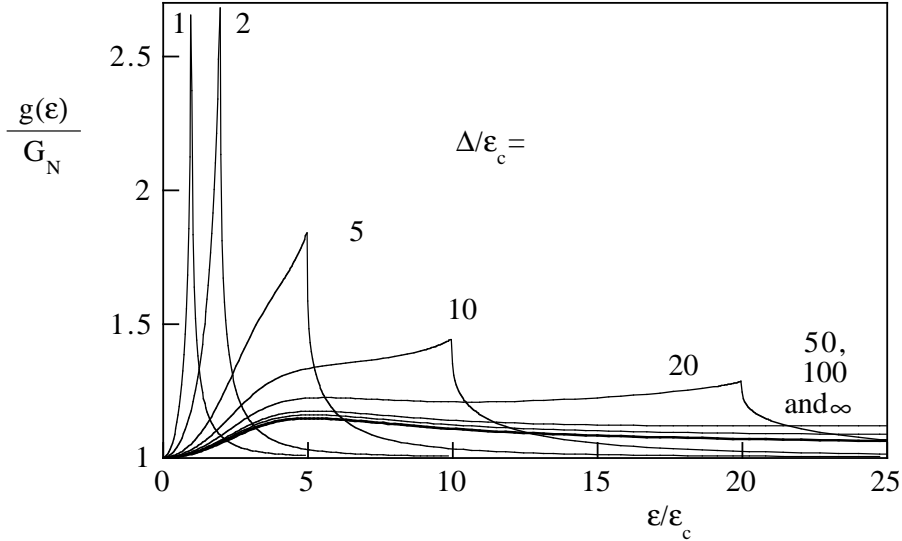}
\caption{Energy dependence of the spectral conductance for various 
values of the superconducting gap in S. The curves are calculated for 
different values of the ratio $\Delta/\epsilon_c$ of the superconducting gap 
$\Delta$ over the Thouless energy $\epsilon_c$. The barrier transparency is 
assumed to be 1. The phase-breaking length is considered as infinite.}
\label{Gap}
\end{figure}

\subsubsection{The superconductor gap}

If both the gap in S and the phase-breaking length are considered 
as infinite, the boundary condition for the complex angle 
at the N-S interface is as simple as : $\theta = i \pi/2$. This 
assumption is valid at low bias and in a restricted temperature 
range, i.e. not too close to the critical temperature of the 
superconductor. Taking into account the finite value of the S gap, 
the boundary condition for the angle $\theta$ has to be changed to :
\begin{equation}
\theta=\frac{\pi}{2}+ \mbox{$i\,$argth}(\frac{\epsilon}{\Delta})
\end{equation}

Fig. \ref{Gap} shows the calculation results for various values of the 
ratio between the Thouless energy and the S gap. If one considers the 
whole behaviour of the spectral conductance $g(\epsilon)$ as a function 
of the energy $\epsilon$, the infinite gap assumption is valid only in 
the regime $\epsilon_{c} \ll \Delta$.
When the gap becomes of the order of the Thouless energy, 
a peak of the spectral conductance enhancement is observed at 
$\epsilon \simeq \Delta$ in addition to the re-entrance peak at $\epsilon 
\simeq 5\epsilon_{c}$. This additional peak is due to the singularity of the 
density of states in the superconductor at the gap energy. The 
amplitude of this additional peak in the spectral conductance can be 
very large. This peak has been recently discussed in Ref. \cite{Peak_Lambert}. 

At low values of the gap $\Delta < 20 \, \epsilon_{c}$, only one peak 
at $\epsilon \simeq \Delta$ is visible. Let us note that even at 
$\Delta = 20 \, \epsilon_{c}$, the contribution of the 
peak of the spectral conductance near $\Delta$ cannot be neglected. 
This will be confirmed in the experimental discussion.

\section{THE EXPERIMENTAL RESULTS}

\subsection{The sample configuration}

\begin{figure} [t]
\epsfxsize=7 cm
\epsfbox{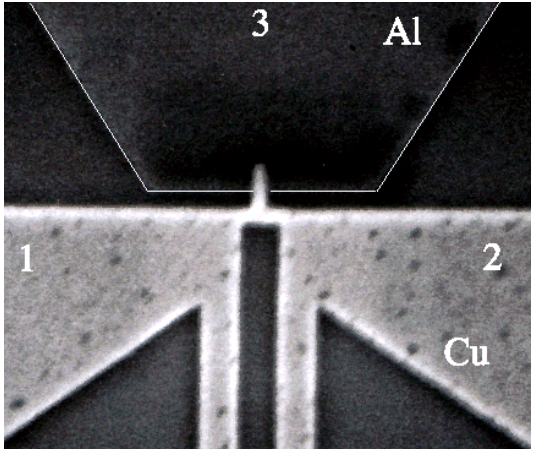}
\caption{Micrograph of the sample. The sample is made of three arms 
of Cu wire joining the two Cu reservoirs "1" and "2" and one Al superconducting 
island "3". Each arm is 200 nm long, 80 nm wide and 50 nm thick. The voltage 
probes of the two Cu reservoirs are visible. The Al island is also 
connected to two measurement probes. For clearity, a thin white line has 
been drawn around of the Al island. Transport measurement were carried out both 
between the two Cu reservoirs ("N-n-N" geometry) and between one Cu reservoir 
and the Al island (N-n-S geometry).}
\label{Photo}
\end{figure}

Fig. \ref{Photo} shows the micrograph of the sample we 
designed for measuring the re-entrance effect. Preliminary 
results from this sample were 
previously published in Ref.\cite{LT}. This T-shaped sample 
is made of three Cu arms joining three wide banks : two are made of 
Cu, one is of Al. The Al island is superconducting below about 
$T_{c}=1.3\,K$. The Cu-Al interface has been carefully prepared in view of 
obtaining a highly transparent interface. 
From the fit to the theory (see below), we derived a maximum  resistance 
value of $2\, \Omega$ for the N-S interface of area about $200 X 
80\,nm^2$. This corresponds  to a length $L_{t}$ of about $80\, nm$, 
which is of the order of the length of the overlapping region between 
the Cu and Al layers.

The conductance can be measured 
between the two Cu reservoirs ("N-n-N") or between one Cu 
reservoir and the Al island ("N-n-S"). With the first geometry, 
the redistribution of current paths in the vicinity of the N-S 
interface\cite{NS_a_2D} due to the superconductivity of Al has a 
reduced effect on the measured conductance. The normal-state 
conductance measured between the 
two Cu reservoirs is $G_{12,N}=0.099\, S$. This gives a 
mean free path $l_{e}$ of $6\, nm$ and a diffusion coefficient $D = v_F 
l_{e}/3=30\, cm^2/s$, using $\rho l_{e}=1.57 \,10^6 \,m.s^{-1}$ for 
Cu. From this data, we can calculate a value for the Thouless energy 
$\epsilon_{c}=12\, \mu eV$ associated with the Cu wire length 
between the two Cu reservoirs. This value is much smaller than the 
gap $\Delta \simeq 190 \, \mu eV$ of Al.

The normal-state conductances of the left (labeled 1) and right (2) arm were measured 
to be $G_{1N}=0.190\, S$ and $G_{2N}=0.206\, S$ respectively. The conductance of the 
upper arm in direct contact with the Al island could not be measured 
directly since the voltage probe was not in close vicinity of the N-S interface. 
From the fit of the experimental results to the theory, we infer 
a value $G_{3N}=0.183\, S$ which is compatible with the sample geometry.
In the following, all the transport measurements were carried out with an ac bias current 
corresponding to a voltage modulation of $2\,\mu eV$. This results in 
a temperature smearing of $23\, mK$.

\subsection{The transport measurements}

\begin{figure} [t]
\epsfxsize=12 cm
\epsfbox{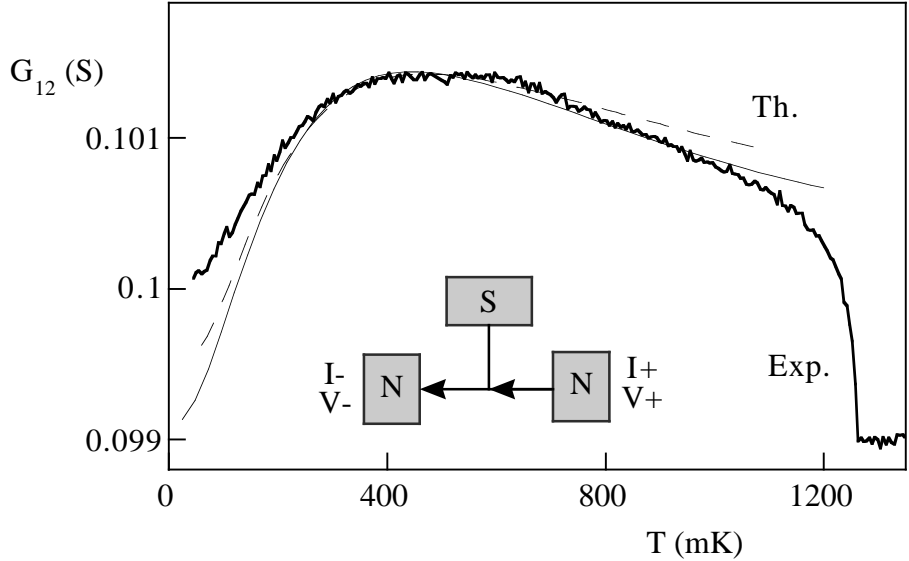}
\caption{Temperature dependence of the "N-n-N" sample 
conductance $G_{12}$ measured between the two N reservoirs at zero 
bias. The experimental curve (Exp.) is shown in parallel with two theoretical 
curves (Th.) calculated with the non-linear Usadel 
equations. The dotted line has been calculated with the assumption of 
an infinite gap in S and the nominal $\epsilon_{c} = 12\, \mu eV$. The full line is 
with $\epsilon_{c} = 15.5\, \mu eV$ and takes into account a constant
value for the gap in the Al superconducting island. The 
phase-breaking length is considered as infinite and the interface 
transparency is taken as equal to 1.}
\label{Reent_T}
\end{figure}

\begin{figure}
\epsfxsize=12 cm
\epsfbox{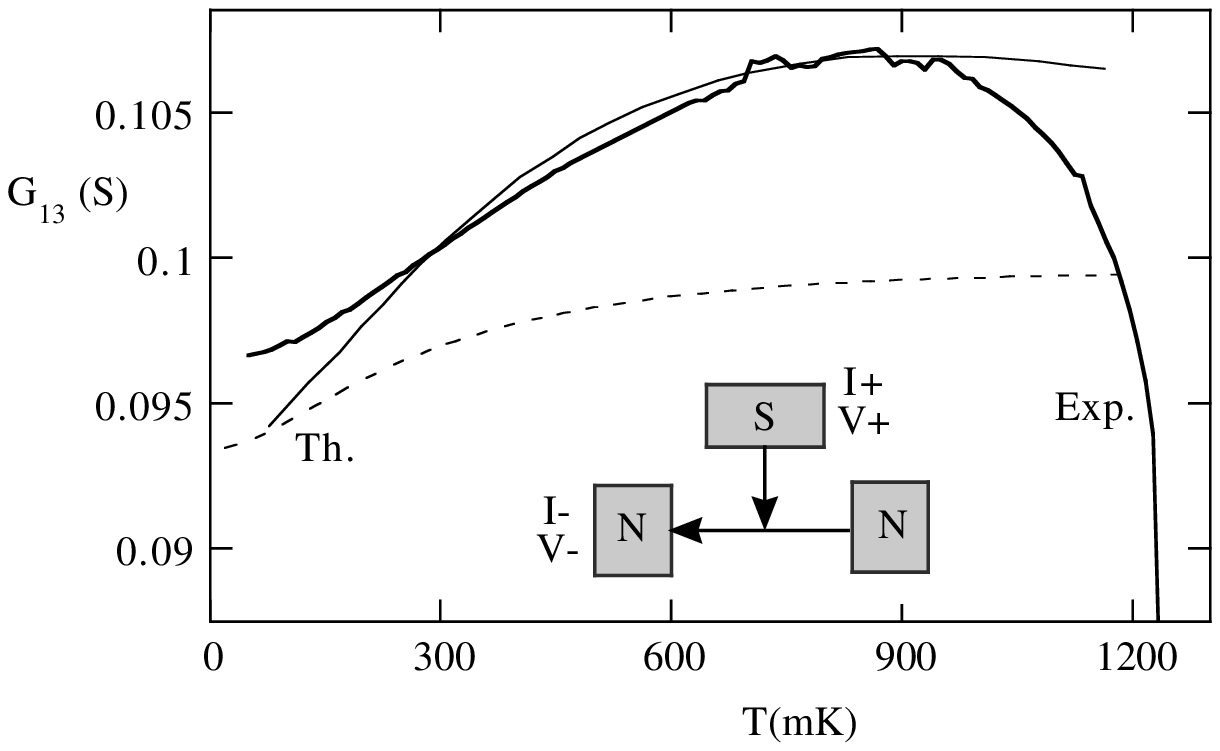}
\caption{Temperature dependence of the "N-n-S" sample 
conductance $G_{13}$ measured between one Cu reservoir and the Al island. 
Same experimental conditions and fit parameters than previous Figure.}
\label{Mes_NS}
\end{figure}

Fig. \ref{Reent_T} shows the temperature dependence of the 
"N-n-N" conductance $G_{12}$. As the temperature decreases below the 
superconducting transition of Al, the conductance first rises. It 
afterwards reaches a maximum near $T = 450\, mK$ and eventually decreases 
down to the lowest temperature ($40\, mK$). This behaviour is 
characteristic of the re-entrance effect. Here, the conductance 
decrease amplitude is comparable to the amplitude of the conductance 
increase at higher energy. Nevertheless, the low-temperature limit of 
the conductance is significantly higher than the normal-state 
conductance.

The "N-n-S" conductance $G_{13}$ measured between the 
Cu reservoir 1 and the Al island is shown in Fig. \ref{Mes_NS}. The 
behaviour is qualitatively similar but with quantitative differences. 
The conductance enhancement is about $10 \% $ between $1.2\, K$ and $40\, 
mK$, instead of $2.5 \%$ for $G_{12}$. In the vicinity of the 
critical temperature of Al, the conductance drops sharply. This 
behaviour is difficult to analyze because the 
voltage drop in Al measured in series is also expected to depend strongly 
on the temperature.

\begin{figure}[t]
\epsfxsize=12 cm
\epsfbox{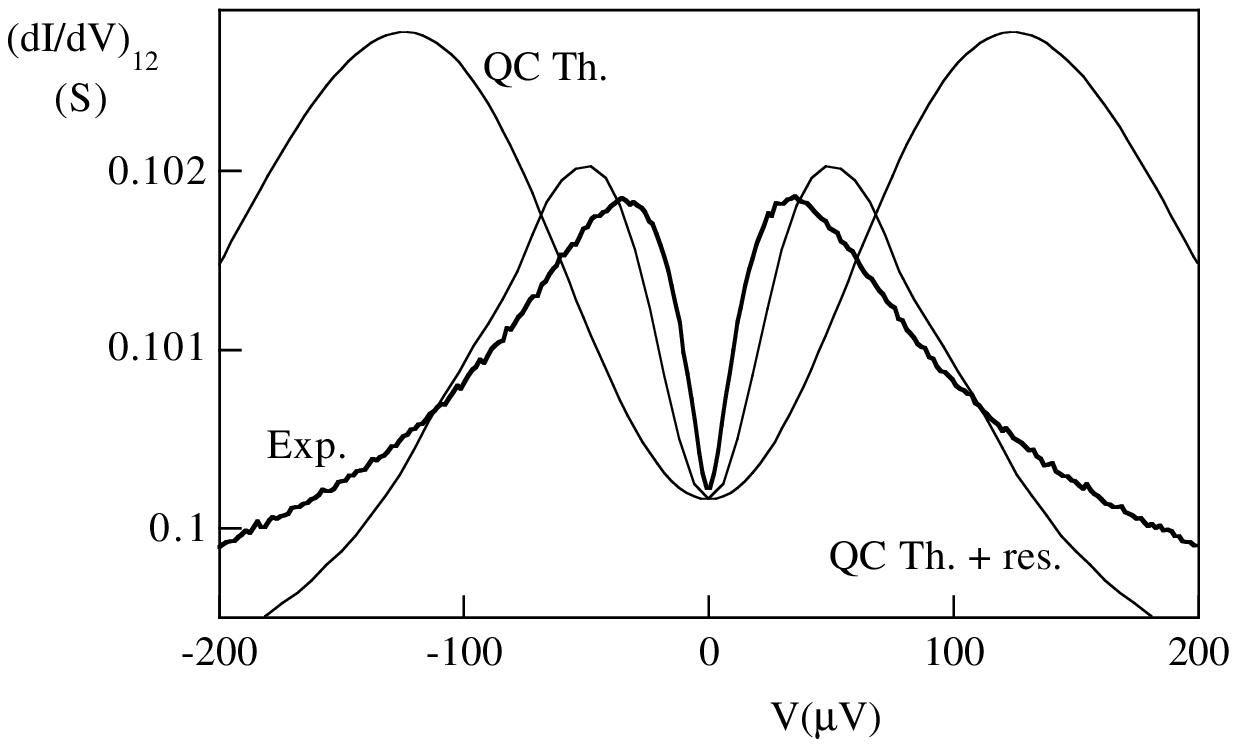}
\caption{Voltage dependence of the differential conductance of the 
same sample in the N-n-N measurement geometry. The theoretical curve labeled (QC Th.) is the 
result of the calculation assuming perfect Cu reservoirs, i.e. with a 
electron Fermi distribution at the phonon bath temperature. The curve 
labeled (QC Th. + res.) includes the additional effect of heating of 
the reservoirs by the bias current. The fit parameters are the same as 
in the two previous figures.}
\label{Reent_V}
\end{figure}

The bias dependence of the non-linear "N-n-N" conductance was measured at $40 \, mK$ 
with an ac modulation of the bias current superposed to a dc current 
bias. Data shown in Fig. \ref{Reent_V} exhibit a 
behaviour which is very similar to the temperature dependence data, 
both in energy dependence and amplitude. 
The differential conductance is minimum at zero bias voltage.
It shows a maximum at a finite bias voltage 
of about $40\, \mu eV$ and eventually a decrease at large bias.

In summary, the conductance of N-S structure is shown to be 
maximum at finite temperature or bias voltage. These two features 
bring the proof for the re-entrance of the spectral conductance since 
both experiments probe the spectral conductance. In the zero bias and 
variable temperature 
experiment, energy is driven by the bath temperature ($\epsilon \simeq 
k_{B}T$). In the very-low-temperature and variable bias experiment, 
energy is driven by the chemical potential of the reservoirs.

The re-entrance of the metallic conductance in a mesoscopic proximity 
superconductor has been previously observed in other 
metallic samples\cite{PrlCharlat,Petr_1998} and in semiconductor 
structures.\cite{Poirier}
The reentrance of the magnetoresistance oscillations has also been 
tracked as a function of the bias voltage in 
semiconductor-superconductor structures, the semiconductor 
being a two-dimensional electron gas.\cite{Hartog} 

\section{DISCUSSION}

\subsection{The detailed spectral conductances}

\begin{figure}[t]
\epsfxsize=11 cm
\epsfbox{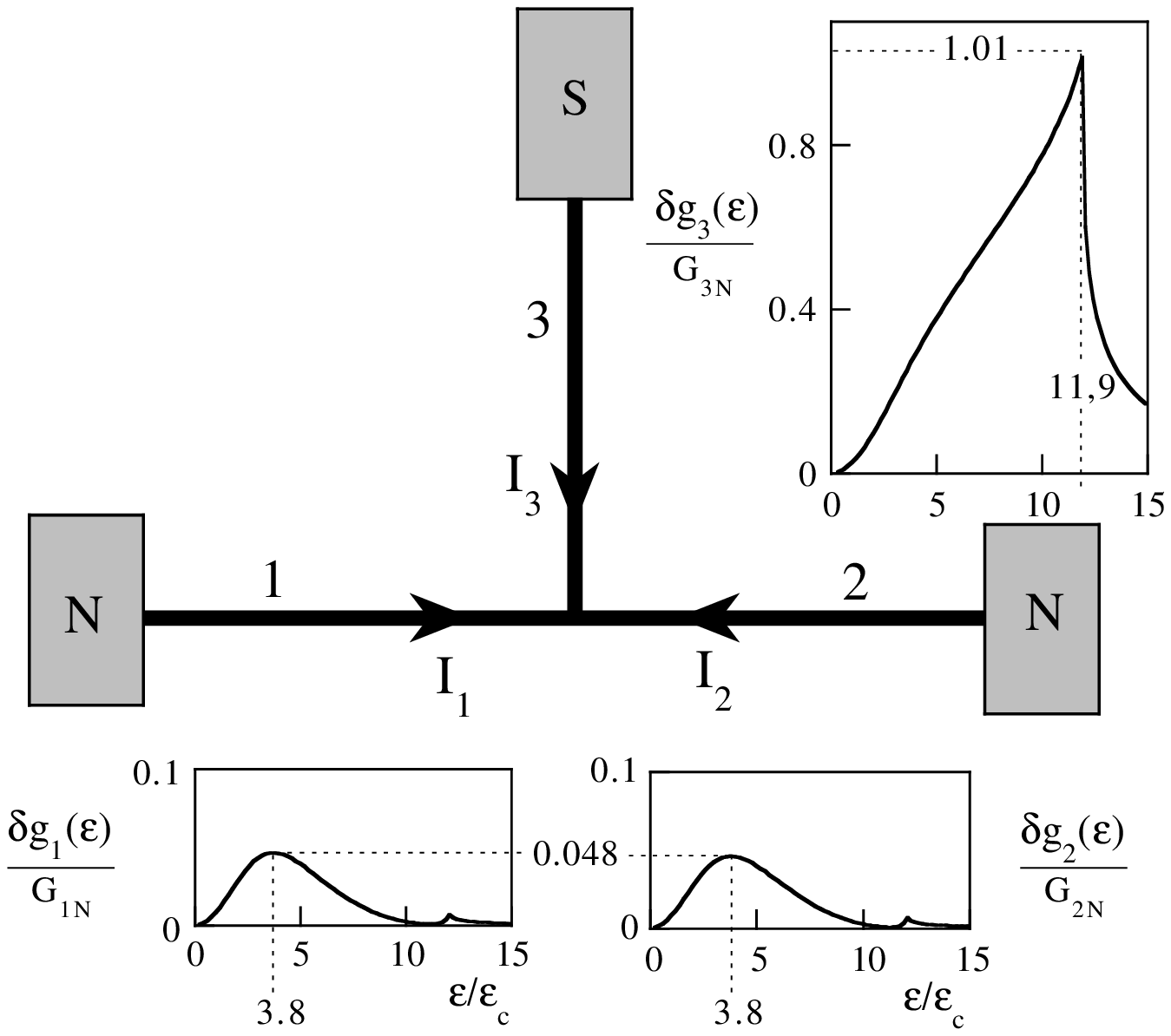}
\caption{Detail of the calculated spectral conductances of each of 
three arms constituing the T-shaped sample. The assumed normal-state 
conductances are $G_{1N}=0.190\, S$, $G_{2N}=0.206\, S$ and $G_{3N}=0.183\, S$. 
The interface transparency is equal to 1 and the phase-breaking length 
is considered as infinite. The gap $\Delta$ of S is equal to 12 times 
the Thouless energy.}
\label{Detail}
\end{figure}

We used the quasiclassical theory to quantitatively describe the 
experimental results. First, the non-linear Usadel equation 
(Eq. \ref{Usadel}) was numerically solved in order to obtain the 
function $\theta_{2}(x,\epsilon)$.
Here we used ideal boundary conditions : $\theta=0$ at the N reservoir, 
perfect N-S interface with a step-like energy gap (no self-consistency), 
continuity for $\theta$ and $\delta_{x}\theta$ at the central node. 
The three spectral conductances 
$g_{1}(\epsilon)$, $g_{2}(\epsilon)$ and $g_{3}(\epsilon)$ are 
afterwards derived from the $\theta_{2}$ function.

Fig. \ref{Detail} shows the three theoretical spectral 
conductances of the three Cu arms of the sample. The energy 
unit is the Thouless energy $\epsilon_{c}=\hbar D/L^2$ 
with the length of each branch assumed to be equal to $L/2$. 
The Thouless energy was taken as a fit parameter. In the following, we will see that the 
agreement between theory and experiment is better with a Thouless 
energy of $15.5\, \mu eV$. This value is indeed 
compatible with our experimental uncertainty on the physical 
dimensions of the sample. The gap of the superconductor Al is $\Delta \simeq 190\, \mu eV$.

From Fig. \ref{Detail}, the spectral conductances of wires 1 and 2 
are maximum at $3.8 \, \epsilon_{c}$. The amplitude of the conductance 
enhancement (here $4.8 \, \%$) is significantly smaller than in Fig. 
\ref{Lin} because the S island is  
separated from the wires 1 and 2. A slight local maximum at higher 
energy is related to the gap of Al. The conductance of the arm 3 is much 
more affected by the finite amplitude of Al gap. This is because 
Andreev pairs with an energy close to the gap remain coherent only 
very close to the N-S interface. The maximum spectral conductance 
of arm 3 is found slightly below the gap of Al at energy 
$\epsilon = 11 \,\epsilon_{c} \simeq \Delta$. The amplitude of this 
peak ($\delta g_{3}/G_{3N} \simeq 100 \%$) is very large. This is 
remarkable, since the gap of S only appears in the boundary condition 
at the N-S interface.

\subsection{From the spectral conductances to the I-V 
characteristics}

Following Eq. \ref{matrice}, a matrix relation makes the link between 
the spectral current in each of the three wires and the out-of-equilibrium 
function $f$ at the nodes of the sample. In our 3-branches circuit, we have :
\begin{equation}
\label{Kirch}
\left\{\begin{array}{l} \displaystyle i_{1}=g_{1} \, [f_{1}-f_{N}] 
 \\ i_{2}=g_{2} \, [f_{2}-f_{N}]   \\
i_{3}=-g_{3} \, f_{N}
\end{array} \right.
\end{equation}	 
where all the quantities are a function of $\epsilon$. Here $f_{1}$, 
$f_{2}$ and $f_{N}$ are the out-of-equilibrium part $f$ of the energy 
distribution functions in reservoirs 1, 2 and at the central node. 
Since the node 3 is the superconducting island, its 
out-of-equilibrium distribution function $f_{3}$ is zero.
The superconducting island 3 is the reference for the voltage. 
In a N-metal reservoir at voltage $V$ at perfect thermal equilibrium at 
temperature $T$, $f$ is of the form :\cite{Volkov}
\begin{equation}
f(\epsilon)=\frac{1}{2}[\tanh\{\frac{\epsilon + eV}{2k_{B}T}\}-
\tanh\{\frac{\epsilon - eV}{2k_{B}T}\}]
\end{equation}	 
According to the Kirchoff law $i_{1}+ i_{2}+ i_{3}=0$ at each 
energy and following Eq. \ref{Kirch}, $f_{N}$ is a 
linear combination of the reservoirs distribution functions :
\begin{equation}
f_{N}=\frac{g_{1} \, f_{1}+g_{2} \, f_{2}}{g_{1}+g_{2}+g_{3}}
\label{dist_fct}
\end{equation}	 
where again all the quantities depend on the energy $\epsilon$.

The I-V characteristics were calculated by integration of the 
equations \ref{Kirch} taken into account the boundary conditions. 
In the two experimental configurations, they are :
\begin{equation}
\begin{array}{lcccc} \displaystyle \mbox{N-n-N}\,\displaystyle: & I_{1}=-I, & I_{2}=I, 
	& I_{3}=0, & V=V_{1}-V_{2}  \\
\displaystyle \mbox{N-n-S} \displaystyle \,: & I_{1}=-I, & I_{2}=0, & I_{3}=-I, & V=V_{1}\end{array}
\end{equation}
 
Let us first consider the zero-bias temperature dependence of the
conductances. In both Fig. \ref{Reent_T} and \ref{Mes_NS}, 
calculated curves are shown in comparison of the experimental one. 
Calculated curves are shown in the two cases of an infinite 
gap taking $\epsilon_{c} = 12\, \mu eV$ and with a finite 
gap $\Delta = 12 \, \epsilon_{c}=186 \, \mu eV$ while 
$\epsilon_{c} = 15.5\, \mu eV$. In the last case, the Thouless energy 
corresponds to a Thouless temperature $\epsilon_{c}/k_{B}=180\, mK$.  
The ratio $\Delta/\epsilon_{c}$ is about 12.

The agreement between experiment and theory is satisfactory 
in the N-n-N geometry in both cases. In the N-n-S geometry, the 
infinite gap assumption does not provide a good description of the 
data. The agreement is good in the case
of a finite gap and $\epsilon_{c} = 15.5\, \mu eV$. 
The fit is not very accurate near the critical temperature $T_{c}$. 
This is because our assumption of a constant value for the 
superconducting gap is not valid anymore in this region.

From these fits, we conclude that the experimental temperature 
dependence data are well described by the theory. The value of the 
physical parameters introduced in the calculation are within the measurement 
uncertainty. The effect of the finite gap is non-negligeable, especially 
in the N-n-S geometry where the conductance of arm 3 contributes. 
This is because the pair amplitude 
$F(\epsilon,x)$ is very large at energies close to the superconductor 
gap. The Andreev pairs with this value of energy remain coherent only very close to the 
N-S interface.

At zero energy 
and zero temperature, one should get the normal-state value of the 
conductance. This is not seen in the experiment. This may 
be an effect of the interactions in the normal metal.\cite{Nazarov96} 
However, our experiment cannot bring a definitive 
answer.\cite{These_Pierre}

Let us now consider the finite bias data. As expected the re-entrance 
effect, i.e. a conductance peak at energy close to the Thouless 
energy, is observed in both the calculated and measured data. However, there is no
quantitative agreement, as the peak position is at significantly lower 
energy than expected. 

\subsection{The role of electron energy distribution in the reservoirs}

In order to obtain a better description of the bias dependence experiment, one has 
to consider the broadening of the energy distribution function in 
the Cu reservoirs by the injected current. We choose to describe this 
heating effect by an effective temperature $T_{eff}$ of the reservoir. 
This effective temperature is different from the phonon temperature 
and depends on the bias current which is injected in the reservoir.

Two different physical effects may be the cause for a heating of the 
reservoirs. First, the chemical potential may be ill-defined in the reservoirs. If 
the reservoir resistance per unit length $\rho_{L}$ is not sufficiently small, a 
significant potential drop $\rho_{L} L_{in} I$ appears in the reservoir, where 
$L_{in}$ is the inelastic mean free path. The electron energy distribution 
will be close to a Fermi distribution with an effective temperature 
$T_{eff}$. If the inelastic mean free path is temperature-independent, 
then the effective temperature behaves as $T_{eff} \propto I$.

A second possibility is the Joule effect induced in the reservoirs 
due to the bias current passing through. This heating power will be 
evacuated by the phonons. The phonon temperature $T$ will be again different 
from the effective electron temperature $T_{eff}$.
Let us first consider the dissipation due to local Joule heating 
throughout the sample. The effective temperature then expresses as 
: $T_{eff} \propto I^{2/5}$ in the limit where $T_{eff} >> T$.\cite{Roukes}
In the pure mesoscopic case where the dissipation occurs only in the 
reservoirs, the Joule effect due to the sample resistance will be evacuated in 
the reservoirs. The reservoir temperature becomes therefore 
non-uniform and the diffusion equation for heat has to be solved. 
Again in the limit where $T_{eff} >> T$, the exponent of the 
previous power-law will be changed into $4/7$.

The current modulation of the sample creates a modulation of the 
effective temperature $T_{eff}$. Then, the
derivative of the electron energy distribution function writes :
\begin{equation}
\frac{df}{dV}= e \frac{df}{d \epsilon}+\frac{d f}{d T_{eff}} 
\frac{d T_{eff}}{d V}
\end{equation}
This introduces an additional term to the usual integration of the 
spectral conductance over a thermal window, namely a convolution of 
the spectral conductance with the derivative of a peak function. 
As a consequence, the usual term of the conductance peak is shifted 
in the experiment compared to the case of ideal reservoirs.

We described the energy distribution of the electrons in 
the reservoirs by a Fermi distribution function with a 
current-dependent effective temperature $T_{eff}$. The effective 
temperature follows a power law : $T_{eff} = C 
I^{\alpha}$, with $C$ being a constant common for all the data.
We found a better agreement by choosing 
$\alpha =1$ as compared to the other possible exponents. The 
ajusted value of the constant $C$ gives an estimate for the inelastic 
mean free path $L_{in}\simeq 4 \mu m$.
The result of the calculation is shown in parallel with the 
experimental curve in Fig. \ref{Reent_V}. The agreement between 
experiment and theory is quite improved. The position and amplitude of 
the conductance peak are well described.

\subsection{A direct test of the non-equilibrium distribution function}

\begin{figure}
\epsfxsize=12 cm
\epsfbox{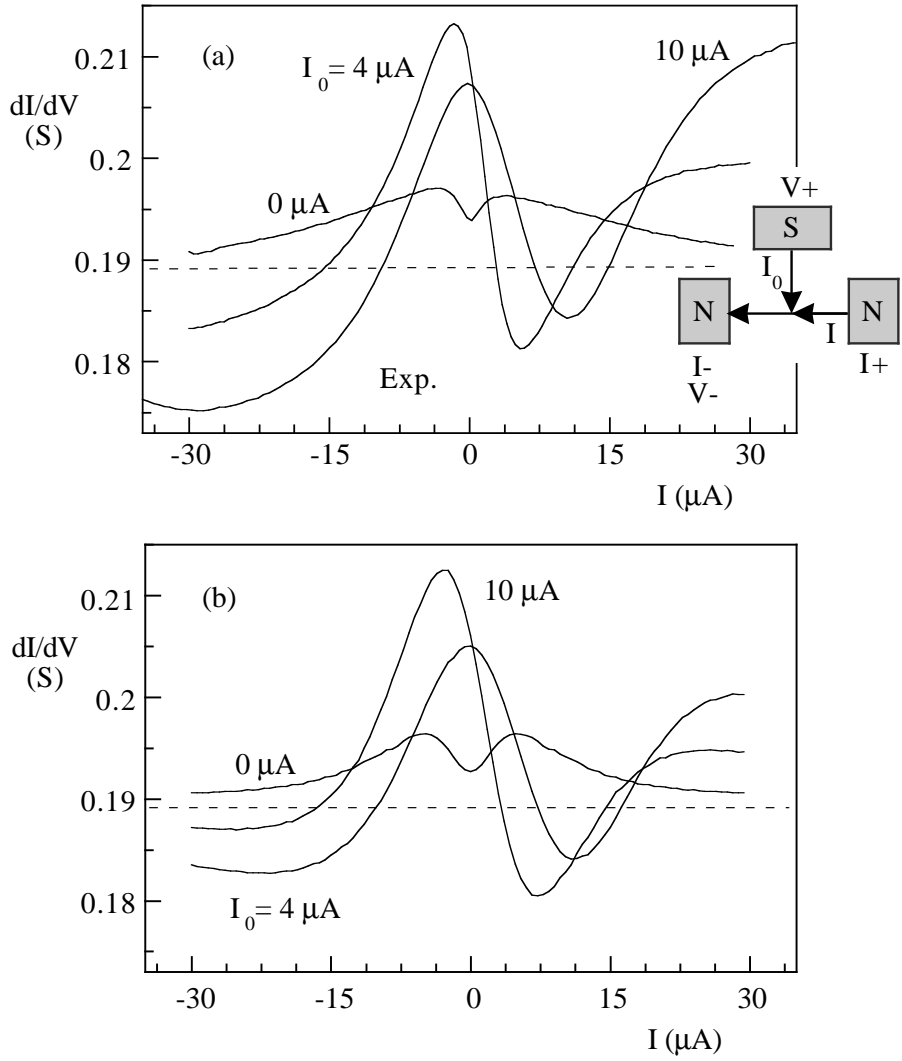}
\caption{(a) : Measured differential conductance of the left arm of the 
sample as a function of the bias current for various values of the 
hot electron injection current $I_0$. Temperature is $100 \, mK$. (b) : Calculated 
differential conductance at $T= 100 \,mK$ of the left arm of the 
sample as a function of the bias current for various values of the 
injection current $I_0$. In both curves sets, the dashed 
line indicates the normal-state conductance.}
\label{Inj}
\end{figure}

Here we describe a complementary experiment designed to test the 
assumption of independent energy channels. The strong point is that 
along the wire, the electron distribution function is strongly 
non-equilibrium, see e.g. Eq. \ref{dist_fct}. Indeed, it consists of 
a superposition of step functions centered at the chemical 
potential of the respective normal reservoirs. The electron energy 
distribution has been directly measured in Ref. \cite{Dist_Gueron}.

The concept is to use the part 3 of the sample as a probe for the 
distribution function at the node of the sample. The part 3 of the 
sample has a strongly energy-dependent conductance, like a N-I-S 
tunnel junction. We injected a dc 
current $I_{o}$ from the Al probe (3) to the left Cu probe (1) while we 
measured the resistance of the left arm (1) of Cu. This was made by 
biasing with a low-frequency ac current between the two Cu probes (1 and 
2) and measuring the voltage drop between the Al probe (3) and the left Cu probe 
(1), see Fig. \ref{Inj}a inset. With a zero injection current, the data 
show the same behaviour as the 
full Cu wire, see Fig. \ref{Reent_V}. With a non-zero injection current, the differential 
conductance becomes asymmetric in respect of the bias current. Peaks 
arise, which shift with varying the injection current from $4$ to $10\, \mu A$. 

In this configuration, the current in the three branches are 
$I_{1}=-(I+I_{0})$, $I_{2}=I$, $I_{3}=I_{0}$. As already stressed, the 
part 3 has the strongest energy dependence of the spectral 
conductance. For simplicity, our qualitative discussion 
will neglect the proximity-induced 
enhancement of the conductance of branches 1 and 2 and keep the 
contribution of the branch 3 only. We will 
restrict our qualitative discussion to the zero-temperature limit. By 
integrating and inverting Eq. \ref{Kirch}, we obtain :
\begin{equation}
\left\{\begin{array}{l} \displaystyle 
V_{1}=-\frac{I}{G_{1N}}-[
\frac{1}{G_{1N}}+\frac{1}{G_{3N}}] I_{0}+\delta V_{N} 
\\
\displaystyle  V_{2}=\frac{I}{G_{2N}}-\frac{I_{0}}{G_{3N}}+\delta V_{N} 
\\
\displaystyle V_{N}=-\frac{I_{0}}{G_{3N}}+\delta V_{N} 
\end{array} \right.
\label{tensions}
\end{equation}
The non-linear contribution $\delta V_{N}$ contains the effect of the 
energy-dependent conductance enhancement $\delta g_{3}(\epsilon)$ :
\begin{equation}
\delta V_{N}=-\int^{\infty}_{-\infty}{\frac{\delta g_{3}(\epsilon)}
{G_{3N}} f_{N}(\epsilon) d\epsilon}
\end{equation}	 

In our experiment, we measure the differential resistance $dV_{1}/dI$ 
versus $I$ with $I_{0}$ kept constant by an independent floating 
source. In the normal-state $1/G_{1N}$ is measured. Below 
$T_{c}$, the non-linear term brings a significant contribution. It measures 
directly the important features of the distribution function at the 
central node.
\begin{equation}
\frac{dV_{1}}{dI}|_{I_{0}}=-\frac{1}{G_{1N}}-\int^{\infty}_{-\infty}{\frac{\delta 
g_{3}(\epsilon)}{G_{3N}} \frac{d f_{N}(\epsilon)}{d I}}
\end{equation}

At very low temperature, the derivative of the distribution function 
of the reservoirs $V_{1}$ and $V_{2}$ should tend to $\delta-$functions 
centered at the chemical potentials of the reservoirs. In this 
hypothesis, it is straightforward to transform the convolution 
product into a combination of $\delta g_{3}(eV_{1})$ and $\delta 
g_{3}(eV_{2})$ using Eq. \ref{dist_fct}, \ref{Kirch} and \ref{tensions}. 
The final formula for the conductance $dI/dV$ (here $V=-V_{1}$) is :
\begin{equation}
\frac{dI}{dV_{1}}|_{I_{0}}=G_{1N}[\,1+\frac{G_{1N}}{G_{N}\,G_{3N}}\,\{\delta 
g_{3}(eV_{2})-\delta g_{3}(eV_{1})\}\,]
\label{Inject_th}
\end{equation}
with $G_{N} = G_{1N}+G_{2N}+G_{3N}$.

From Eq. \ref{tensions}, we notice that both $V_{1}$ and $V_{2}$ are tuned by the independent 
driving currents $I$ and $I_{0}$. They vanish respectively for 
$I^-=-I_{0}(1+\,G_{1N}/G_{3N})$ and 
$I^+=I_{0}\,G_{3N}/G_{2N}$. Referring to Fig. \ref{Detail}, the 
strong minimum of $\delta g_{3}(\epsilon)$ at $\epsilon =0$ 
show up as a minimum of $\frac{dI}{dV_{1}}$ at $I^+$ where $V_{2}=0$ 
and a maximum at $I^-$ 
where $V_{1}=0$. This is precisely the origin of the strong assymetric 
behaviour observed in the experimental curve of Fig. \ref{Inj}.
The asymmetry is present because of the distribution 
function $f_{eN}$ at the central node exhibits sharp steps at the 
chemical potentials of the reservoirs $eV_{1}$ and $eV_{2}$. With 
this experiment, we have proven that the energy distribution of the 
electrons at the node in the middle of our sample is truly 
out-of-equilibrium. In brief, we are indeed in a mesoscopic regime.

In order to achieve a quantitative comparison between theory and experiment, we 
solved the full matrix expression Eq. \ref{matrice} which makes no 
simplification on $\delta g_{1}(\epsilon)$ and $\delta 
g_{2}(\epsilon)$.We included the heating effect in the reservoirs in 
the same way than in the re-entrant conductance data. The data of 
Fig. \ref{Inj} are well described by using the same constant 
$C$ and exponent 1 as in the Fig. \ref{Reent_T} and \ref{Reent_V}.
As expected, the two extrema occur at values of $I$ proportional to 
$I_{0}$. When $I_{0}=0$, this contribution is absent. Only the smaller 
symmetric contribution of $\delta g_{1}(\epsilon)$ remains. Our 
previous analysis leading to Eq. \ref{Inject_th} ignores this 
contribution but highlights the main contribution due to $\delta 
g_{3}(\epsilon)$. We also studied an alternate geometry with the inversion of the 
currents $I$ and $I_{0}$.\cite{These_Pierre} In this case, the same 
two terms $\delta g_{3}(eV_{1})$ 
and $\delta g_{3}(eV_{2})$ are added instead of being subtracted. 
The experimental data (not shown) were again consistently fitted by 
the calculation with the same physical parameters.

\section{CONCLUSION}

In conclusion, we presented a thorough study of the spectral conductance of a 
mesoscopic normal metal in contact with a superconductor. The spectral 
conductance exhibits a re-entrance effect at zero energy, i.e. the 
zero-energy spectral conductance coincides with the normal state 
conductance. From the theoretical calculations, we observe that 
the spectral conductance is sensitive to the absolute value of the 
many physical parameters including : phase-breaking length, gap of the 
superconductor and interface transparency.

We performed experimental measurement of the reentrance effect as a 
function of both bias voltage and temperature. The experimental data is 
well described by the quasiclassical theory. Nevertheless, 
the description of the non-linear conductance data requires taking 
into account of the heating of the N reservoirs by the bias current.
Compared to a previous study,\cite{Petr_1998} we were able to 
describe quantitatively the reentrant 
resistance with  the quasiclassical theory. Thanks to the 
well-controlled geometry of the normal-metal reservoirs, no scaling 
factor was necessary.

The large energy-sensitivity of the conductance enhancement by the 
proximity effect can be used to probe directly the energy distribution 
function without using a tunnel junction. Thus we were able to measure 
the conductance enhancement and test the distribution function in a 
single sample. The whole set of data agree with the 
theoretically-calculated curves with a single set of physical parameters. 
This confirms that most of the behaviour of this sample is understood 
by the theory of the proximity effect in a non-interacting metal.
In contrast, the difference between the conductance measured in the 
zero-temperature limit and the normal-state value raises 
the question of the importance of the effect of interactions. 

We thank F. W. J. Hekking, T. Martin, B. Spivak, A. F. Volkov 
and F. Zhou for fruitful discussions. This work was supported by the 
DRET, the R\'egion Rh\^one-Alpes and the 
TMR contract "Dynamics of superconducting nanocircuits" from EU.


\begin{thebibliography}{9}
\bibitem{Revue_Lambert} C. J. Lambert and R. Raimondi, J. Phys.: 
Condens. Matter {\bf 10}, 901 (1998) and references therein.
\bibitem{Comments} D. C. Ralph and V. Ambegaokar, Comments Cond. 
Mat. {\bf 18}, 249 (1998).
\bibitem{Esteve} D. Est\`{e}ve, in {\it Mesoscopic Electron Transport},
Eds. L.P. Kouwenhoven, G. Sch\"{o}n and L.L. Sohn, Kluwer Academic
Publishers, Dordrecht, The Netherlands (1996).
\bibitem{Mota} P. Visani, A. C. Mota, A. Pollini, Phys. Rev. Lett. {\bf 
65}, 1514 (1990).
\bibitem{Gueron} S. Gu\'eron, H. Pothier, N. O. Birge, D. Est\`eve 
and M. H. Devoret, Phys. Rev. Lett. {\bf 77}, 3025 (1996).
\bibitem{Petrashov} V. T. Petrashov, V. N. Antonov, P. Delsing, and 
T. Claeson, Phys. Rev. Lett. {\bf 70}, 347 (1993); Phys. Rev. Lett. 
{\bf 74}, 5268 (1995).
\bibitem{CourtoisPrl} H. Courtois, Ph. Gandit, D. Mailly, and B. 
Pannetier, Phys. Rev. Lett. {\bf 76}, 130 (1996).
\bibitem{Taka} V.N. Antonov, A.F. Volkov, and H. Takayanagi, 
Europhys. Lett. {\bf 38}, 453 (1997).
\bibitem{BTK} G. E.  Blonder, M. Tinkham and T.M. Klapwijk, Phys.
Rev. B {\bf 25}, 4515 (1982).
\bibitem{Kastalsky}	A. Kastalsky, A. W. Kleinsasser, L. H. Greene, R. Bhat, F. P. 
Milliken, and J. P. Harbison, Phys. Rev. Lett. {\bf 67}, 3026 (1991).
\bibitem{vanWees} B.J. van Wees, P. de Vries, P. Magnee and T.M.
Klapwijk, {\it Phys. Rev. Lett.} {\bf 69}, 510 (1992).
\bibitem{HekkingNazarov} F .W. J. Hekking and Y. Nazarov, Phys.  Rev.
Lett. {\bf 71}, 1625 (1993); Phys. Rev. B {\bf 71}, 6847 (1994).
\bibitem{Lambert} C. Lambert, J. Phys. Condens. Matter {\bf 3}, 
6579 (1991).
\bibitem{Beenakker} C. W. J. Beenakker, Phys. Rev. B {\bf 46}, 12841 
(1992).
\bibitem{PrlCharlat} P. Charlat, H. Courtois, Ph. Gandit, D. Mailly, A. 
Volkov, and B. Pannetier, Phys. Rev. Lett. {\bf 79}, 4950 (1996).
\bibitem{Hartog} S. G. den Hartog, C. M. A. Kapteyn, B. J. van Wees, 
and T. M. Klapwijk, Phys. Rev. Lett. {\bf 76}, 4592 (1996).
\bibitem{Petr_1998} V. T. Petrashov, R. Sh. Shaikhaidorov, P. Delsing, and 
T. Claeson, JETP Lett. {\bf 67}, 513 (1998).
\bibitem{Poirier} W. Poirier, D. Mailly, and M. Sanquer, Phys. Rev. 
Lett. {\bf 79}, 2105 (1997).
\bibitem{Andreev} A. F. Andreev, Sov. Phys. JETP {\bf 19}, 1228 
(1964).
\bibitem{Blom} H. A. Blom, A. Kadigrobov, A. M. Zagoskin, R. I. 
Shekhter, and M. Jonson, Phys. Rev. B {\bf 57}, 9995 (1998).
\bibitem{de_Gennes} G. Deutscher and P.G. de Gennes, in {\it
Superconductivity}, Vol. 2, R.D. Parks, ed. Marcel Dekker, New York
(1969).
\bibitem{Larkin} A. I. Larkin and Yu. N. Ovchinikov, Sov. Phys. JETP 
{\bf 28}, 1200 (1969).
\bibitem{Volkov} A. F. Volkov, A. V. Zaitsev, and T. M. Klapwijk, 
Physica C {\bf 210}, 21 (1993); A. F. Volkov and A. V. Zaitsev, 
Phys. Rev. B {\bf 53}, 9267 (1996); A. V. Zaitsev, JETP Lett. {\bf 
51}, 35 (1990).
\bibitem{Zhou} F. Zhou, B. Spivak, and A. Zyuzin, Phys. Rev. B {\bf 
52}, 4467 (1995).
\bibitem{Nazarov96} Y. V. Nazarov and T. H. Stoof, Phys. Rev. Lett. 
{\bf 76}, 823 (1996).
\bibitem{Wilhelm} A. A. Golubov, F. K. Wilhelm, and A. D. Zaikin, 
Phys. Rev. B {\bf 55}, 1123 (1997).
\bibitem{Yip} S. Yip, Phys. Rev. B {\bf 52}, 15504 (1995).
\bibitem{Volkov-Lambert} A. F. Volkov, N. Allsopp and C. J. Lambert, 
J. Phys. Condens. Matter {\bf 8}, L45 (1996).
\bibitem{Sols} F. Sols and J. Ferrer, Phys. Rev. B {\bf 49}, 15913 
(1994).
\bibitem{Bagwell} P. Bagwell, Phys. Rev. B {\bf 49}, 6841 (1994).
\bibitem{Lesovik} G. B. Lesovik, A. L. Fauch\`ere, and G. Blatter, 
Phys. Rev. B {\bf 55}, 3146 (1997)
\bibitem{These_Pierre} P. Charlat, PhD Thesis, University Joseph 
Fourier, Grenoble (1997).
\bibitem{FS} M. Giroud, H. Courtois, K. Hasselbach, D. Mailly, and B. 
Pannetier, Phys. Rev. B {\bf 58}, R11872 (1998).
\bibitem{Pannetier-Rammal} B. Pannetier, J. Chaussy, and R. Rammal, 
Phys. Scripta {\bf T 13}, 245, (1986).
\bibitem{Pothier} H. Pothier, S. Gu\'eron, D. Est\`eve, and M. H. Devoret,
Phys. Rev. Lett. {\bf 73}, 2488 (1994).
\bibitem{Peak_Lambert} A. F. Volkov, V. V. Pavlovskii, and R. Seviour, 
to appear in Sup. and Microstructures.
\bibitem{LT} P. Charlat, H. Courtois, Ph. Gandit, D. Mailly, A. 
Volkov, and B. Pannetier, Proceedings of the LT21 Conference, Czech. J. 
of Phys. {\bf 46}, S6 3107 
(1996).
\bibitem{NS_a_2D} F. K. Wilhelm, A. D. Zaikin and H. Courtois, Phys. 
Rev. Lett. {\bf 80}, 4289 (1998).
\bibitem{Dist_Gueron} H. Pothier, S. Gu\'eron, N. O. Birge, D. Esteve, 
and M. H. Devoret, Phys. Rev. Lett. {\bf 79}, 3490 (1997).
\bibitem{Roukes} M. L. Roukes, M. R. Freeman, R. S. Germain, R. C. 
Richardson and M. B. Ketchen, Phys. Rev. Lett. {\bf 55}, 422 (1985).
\end{thebibliography}
\end{document}